# Propagation of Spin Waves Through an Interface Between Ferromagnetic and Antiferromagnetic Materials

Oksana Busel[1,*], Oksana Gorobets[1], Yuri Gorobets[1,2]

[1]Faculty of Mathematics and Physics, National Technical University of Ukraine "Igor Sikorsky Kyiv Polytechnic Institute", Prospect Peremohy 37, Kyiv, 03056, Ukraine
[2]Institute of Magnetism NAS and MES of Ukraine, Vernadskiy Av., 36-b, Kyiv, 03142, Ukraine

[*]opbusel@gmail.com

**Abstract:** Boundary conditions for order parameters at an interface between ferromagnetic (FM) and two-sublattice antiferromagnetic (AFM) materials were obtained in the continuous medium approximation similarly to the approach which allows one to take into account the finite thickness of the FM/FM interface, which is much less than spin wave length. Three order parameters are considered inside an interface of finite thickness with the magnetization **M** of FM, magnetizations of both sublattices $\mathbf{M}_1$ and $\mathbf{M}_2$ of AFM. The uniform and non-uniform exchange between all order parameters are taken into account to the interface energy. Using these boundary conditions, the excitation of a surface evanescent spin wave is considered in AFM when the spin wave in FM falls onto this interface. The coefficients and the phases of transmission and reflection of spin wave through the FM/AFM interface are derived.

## 1 Introduction

The understanding of the boundary conditions for the theory of wave propagation in non-uniform media at interfaces between regions with different material properties is of essential importance, such as at an interfaces between two FMs or FM/AFM [1, 2]. Recent years have witnessed a growing interest to properties of antiferromagnets, since they have a plenty of advantages compared to ferromagnets [3, 4].

Since spin waves in AFMs operate coherently in the THz regime [5], which is orders of magnitude faster than the frequency of typical ferromagnetic spin waves the development of novel methods for exciting AFMs at the nanoscale was investigated [6], which contributed to development in the field of antiferromagnetic spintronics.

AFMs are attractive as potential active elements in next-generation spin-transport and memory-storage devices [7-10] therefore the formation and dynamics of spin textures in antiferromagnetic insulators was investigated and an alternative generic geometry for the induction of ultrafast autonomous antiferromagnetic dynamics was proposed [11].

It is important to note that during the theoretical study of dynamics of antiferromagnetic domain walls driven by spin-orbit torques in AFM/heavy metal bilayers [12] was found that the antiferromagnetic domain wall velocity can reach a few kilometers per second and the antiferromagnetic domain wall can therefore serve as a terahertz source [12]. Furthermore, motion of topological solitons in AFMs was researched under the combined action of perturbations such as an external magnetic field and torque-generating electrical current and it was shown that spins of electrons exchange angular momentum with the soliton [13].

The exchange bias (EB) effect in the magnetic systems composed of FM/AFM also causes considerable amount of interest since the FM/AF interface plays a significant role on the EB properties of magnetic multilayers and nanoparticles [14], and the physical properties of such systems have been widely investigated [15–17].

According to the foregoing, possible cooperation between antiferromagnetism and topological properties in both momentum and real spaces is of great interest [18].

In this paper, the most general form of the boundary conditions between FM and two-sublattice AFM were reduced to the inclusion of only energies of uniform and non-uniform exchange between all sublattices [1], and the transit of a surface evanescent spin wave [19] through the FM/AFM interface have been investigated when the spin wave in FM falls onto this interface in conformity with the results of the previous work [1].

## 2 Theoretical Consideration

### 2.1 Abridged general boundary conditions in the interface between FM/AFM

The tasks of this work are: first – to find the abridged general form of the boundary conditions between FM and two-sublattice AFM taking into account the fact that the interface is a composite material with finite thickness δ which is much less than the length of the spin wave $\lambda_{sw}$ including only energies of uniform and non-uniform exchange between all sublattices [1, 2]; and second – to derive the coefficients of transmission and reflection of surface evanescent spin wave through the FM/AFM interface [19].

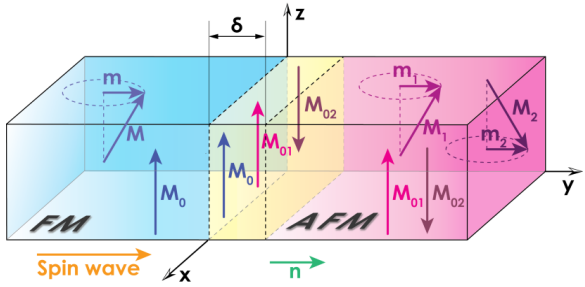

**Fig. 1.** The model shows the schematic image of the system consisting of FM, interface of finite thickness between FM/AFM and two-sublattice AFM and magnetizations in each layer with the small perturbations of order parameters relative to the ground state. The FM layer creates spin wave excitations in the AFM layer.

The normal to the interface of magnets **n** is parallel to the *y*-axis.

The main form of energy taking into account only energies of uniform and non-uniform exchange between all sublattices is following:

$$W = \int w dy$$
$$= \int dy \begin{cases} A_{12}(y)\mathbf{M}_1\mathbf{M}_2 + A_1(y)\mathbf{M}_1\mathbf{M} + A_2(y)\mathbf{M}_2\mathbf{M} + \\ +\alpha'_{12}(y)\left(\frac{\partial \mathbf{M}_1}{\partial y}\right)\left(\frac{\partial \mathbf{M}_2}{\partial y}\right) + \frac{1}{2}\alpha_1(y)\left(\frac{\partial \mathbf{M}_1}{\partial y}\right)^2 + \\ +\frac{1}{2}\alpha_2(y)\left(\frac{\partial \mathbf{M}_2}{\partial y}\right)^2 + \frac{1}{2}\alpha(y)\left(\frac{\partial \mathbf{M}}{\partial y}\right)^2 + \\ +\alpha'_1(y)\left(\frac{\partial \mathbf{M}_1}{\partial y}\right)\left(\frac{\partial \mathbf{M}}{\partial y}\right) + \alpha'_2(y)\left(\frac{\partial \mathbf{M}_2}{\partial y}\right)\left(\frac{\partial \mathbf{M}}{\partial y}\right) \end{cases}. \quad (1)$$

FM is magnetized along the *z*-axis: **M** is parallel to the *z*-axis (as shown in Fig. 1) and the *z*-axis is easy axis in the AFM, where the AFM vector is $\mathbf{L} = (\mathbf{M}_1 - \mathbf{M}_2)$ (in the ground state); $A_{12}(y)$, $\alpha'_{12}(y)$ are uniform and non-uniform exchange magnetic parameters between 1st and 2nd AFM sublattices, respectively; $A_1(y)$, $A_2(y)$, $\alpha'_1(y)$, $\alpha'_2(y)$ are uniform and non-uniform exchange magnetic parameters between FM-1st and FM-2nd AFM sublattices, respectively; $\alpha(y)$, $\alpha_1(y)$, $\alpha_2(y)$ are non-uniform exchange magnetic parameters in the FM layer, 1st and 2nd AFM sublattices, respectively.

The magnetic parameters characterizing FM and AFM materials and material of the interface region in the energy (1) have typical dependency on the *y* coordinate which is illustrated in Fig. 2.

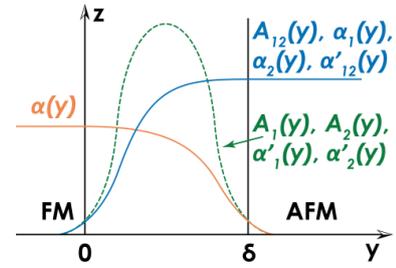

**Fig. 2.** Schematic assumption of the coordinate dependence of the magnetic parameters characterizing FM, two-sublattice AFM, and the interface region in the energy (1) on the *y* coordinate.

Oder parameters **M**, $\mathbf{M}_1$, $\mathbf{M}_2$ are considered as slowly varying functions with $y \in [0,\delta]$, and the coefficients $A_1(y)$, $A_2(y)$, $A_{12}(y)$, $\alpha_1(y)$, $\alpha_2(y)$, $\alpha(y)$, $\alpha'_1(y)$, $\alpha'_2(y)$, $\alpha'_{12}(y)$ as rapidly varying functions (Fig. 2) for deriving boundary conditions taking into account $\delta \ll \lambda_{sw}$.

The values of the magnetic parameters on the interface are presented in Table 1.

**Table 1.** Notations for the character of variation of the magnetic parameters in the vicinity of the interface.

| $\alpha_1(y)\big|_0^\delta \equiv \alpha_1$ | $\alpha'_1(y)\big|_0^\delta \equiv 0$ | $\int_0^\delta A_1(y)dy = A_1$ |
|---|---|---|
| $\alpha_2(y)\big|_0^\delta \equiv \alpha_2$ | $\alpha'_2(y)\big|_0^\delta \equiv 0$ | $\int_0^\delta A_2(y)dy = A_2$ |
| $\alpha(y)\big|_0^\delta \equiv -\alpha$ | $\alpha'_{12}(y)\big|_0^\delta \equiv \alpha'_{12}$ | $\int_0^\delta A_{12}(y)dy = A_{12}$ |

Boundary conditions in vector form for order parameters at an interface between FM and two-sublattice AFM materials have been obtained in the continuous medium approximation taking into account uniform and non-uniform exchange between all sublattices same to the approach in [1]. Boundary conditions in vector form can be written as:

$$\begin{cases} \alpha_1\left[\mathbf{M}_1 \times \frac{\partial \mathbf{M}_1}{\partial y}\right] + \alpha'_{12}\left[\mathbf{M}_1 \times \frac{\partial \mathbf{M}_2}{\partial y}\right] - A_{12}[\mathbf{M}_1 \times \mathbf{M}_2] - \\ \qquad\qquad - A_1[\mathbf{M}_1 \times \mathbf{M}] = 0 \\ \alpha_2\left[\mathbf{M}_2 \times \frac{\partial \mathbf{M}_2}{\partial y}\right] + \alpha'_{12}\left[\mathbf{M}_2 \times \frac{\partial \mathbf{M}_1}{\partial y}\right] - A_{12}[\mathbf{M}_2 \times \mathbf{M}_1] - \\ \qquad\qquad - A_2[\mathbf{M}_2 \times \mathbf{M}] = 0 \\ \alpha\left[\mathbf{M} \times \frac{\partial \mathbf{M}}{\partial y}\right] + A_1[\mathbf{M} \times \mathbf{M}_1] + A_2[\mathbf{M} \times \mathbf{M}_2] = 0 \end{cases}. \quad (2)$$

For example, the conditions (2) have been linearized taking into account the ground states of magnetization of FM, AFM and the interface, considering the small perturbations of order parameters relative to the ground state as following:

$$\mathbf{M} = M_0\mathbf{e}_z + \mathbf{m}, \; \mathbf{m} = m_x\mathbf{e}_x + m_y\mathbf{e}_y$$
$$\mathbf{M}_1 = M_{01}\mathbf{e}_z + \mathbf{m}_1, \; \mathbf{m}_1 = m_{1x}\mathbf{e}_x + m_{1y}\mathbf{e}_y \quad , \quad (3)$$
$$\mathbf{M}_2 = M_{02}\mathbf{e}_z + \mathbf{m}_2, \; \mathbf{m}_2 = m_{2x}\mathbf{e}_x + m_{2y}\mathbf{e}_y$$

where

$$\begin{aligned}|\mathbf{m}| &\ll M_0 \\ |\mathbf{m}_1| &\ll M_{01}, \\ |\mathbf{m}_2| &\ll M_{02}\end{aligned} \quad (4)$$

where $\mathbf{m}$ is deviation magnetization of FM, $\mathbf{m}_1$, $\mathbf{m}_2$ are deviation magnetization of two-sublattices AFM, $M_0$ is the projection of the magnetization of FM, $M_{01}$ and $M_{02}$ are projections of the magnetizations of the first and second sublattices respectively of the AFM to the z-axis in the ground state.

For convenience the next notations are used:

$$x, y = n, \; \frac{M_{02}}{M_{01}} = \gamma = -1, \; \frac{M_0}{M_{01}} = \gamma_0, \; \frac{M_0}{M_{02}} = \frac{\gamma_0}{\gamma} = -\gamma_0.$$

The linearized boundary conditions have the following form:

$$\begin{cases} \alpha_1 \dfrac{\partial m_{1n}}{\partial y} + \alpha'_{12}\dfrac{\partial m_{2n}}{\partial y} - A_{12}(m_{1n} + m_{2n}) - \\ \qquad\qquad\qquad\qquad -A_1(m_n - \gamma_0 m_{1n}) = 0 \\ \alpha_2 \dfrac{\partial m_{2n}}{\partial y} + \alpha'_{12}\dfrac{\partial m_{1n}}{\partial y} - A_{12}(m_{1n} + m_{2n}) - \\ \qquad\qquad\qquad\qquad -A_2(m_n + \gamma_0 m_{2n}) = 0 \\ \alpha \dfrac{\partial m_n}{\partial y} - A_1\left(\dfrac{m_n}{\gamma_0} - m_{1n}\right) + A_2\left(\dfrac{m_n}{\gamma_0} + m_{2n}\right) = 0 \end{cases} \quad (5)$$

**2.2 Propagation of a surface evanescent spin wave through the interface between FM/AFM:**

The excitation of a surface evanescent spin wave has been considered in AFM when spin wave in FM falls onto this interface as shows on the Fig. 3.

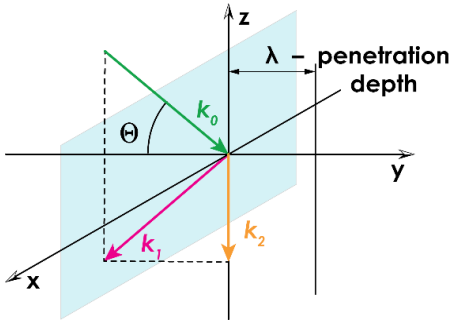

**Fig. 3.** Schematic illustration of the excitation of a surface evanescent spin wave in AFM when spin wave in FM falls onto this interface, where the incident wave vector is $\mathbf{k}_0 = (0, k_{0y}, -k_{0z})$, the reflected wave vector is $\mathbf{k}_1 = (0, -k_{1y}, -k_{1z})$ and the transmitted wave vector is $\mathbf{k}_2 = (0, 0, -k_{2z})$, and $\theta$ - angle between the wave vector of incident wave and y-axis. From Fig. 3, it is clear that $k_{0y} = -k_{1y}$ and $k_{0z} = k_{1z} = k_{2z}$.

Then angle between the wave vector of incident wave and y-axis can be expressed in terms of the wave vector components as follows:

$$k_0 = \sqrt{k_{0y}^2 + k_{0z}^2}, \; \cos\theta = \frac{k_{0y}}{\sqrt{k_{0y}^2 + k_{0z}^2}} \; . \quad (6)$$

The evanescent spin wave can be written in a following form:

$$\begin{aligned} m_n &= A_0 \exp(i(k_{0y}y + k_{0z}z - \omega t)) + \\ &\quad + R\exp(i(-k_{0y}y + k_{0z}z - \omega t + \varphi_1)) \\ m_{1n} &= A\exp(-y/\lambda)\exp(i(k_{2z}z - \omega t + \tilde{\varphi})), \\ m_{2n} &= B\exp(-y/\lambda)\exp(i(k_{2z}z - \omega t + \tilde{\tilde{\varphi}})) \end{aligned} \quad (7)$$

where $A_0$ – amplitude and $\omega$ – frequency of incident spin wave which are defined.

$A$, $B$, $R$, $\tilde{\varphi}$, $\tilde{\tilde{\varphi}}$ and $\varphi_1$ should be found, where $A$, $B$ are amplitudes of transmitted spin wave (two amplitudes is a result of consideration of two-sublattices AFM); $\tilde{\varphi}$, $\tilde{\tilde{\varphi}}$ are phases of transmitted spin wave; $R$ and $\varphi_1$ are amplitude and phase of oscillations of a reflected spin wave, respectively.

After substituting Eqs. (7) in the linearized boundary conditions (5) and simplify the system of equations can be written as:

$$\begin{cases}\dfrac{1}{\lambda}\alpha_1 A\cos\tilde{\varphi} + \dfrac{1}{\lambda}\alpha'_{12}B\cos\tilde{\tilde{\varphi}} + A_{12}A\cos\tilde{\varphi} + A_{12}B\cos\tilde{\tilde{\varphi}} - \\ \qquad\qquad -A_1 A\gamma_0\cos\tilde{\varphi} + A_1 A_0 + RA_1\cos\varphi_1 = 0 \\ \dfrac{1}{\lambda}\alpha_2 B\cos\tilde{\tilde{\varphi}} + \dfrac{1}{\lambda}\alpha'_{12}A\cos\tilde{\varphi} + A_{12}A\cos\tilde{\varphi} + A_{12}B\cos\tilde{\tilde{\varphi}} - \\ \qquad\qquad -A_2 B\gamma_0\cos\tilde{\tilde{\varphi}} + A_2 A_0 + RA_2\cos\varphi_1 = 0 \\ k_{0y}\alpha R\sin\varphi_1 - \dfrac{1}{\gamma_0}A_0 A_1 - \dfrac{1}{\gamma_0}A_1 R\cos\varphi_1 + AA_1\cos\tilde{\varphi} + \\ \qquad +\dfrac{1}{\gamma_0}A_0 A_2 + \dfrac{1}{\gamma_0}A_2 R\cos\varphi_1 + BA_2\cos\tilde{\tilde{\varphi}} = 0 \\ \dfrac{1}{\lambda}\alpha_1 A\sin\tilde{\varphi} + \dfrac{1}{\lambda}\alpha'_{12}B\sin\tilde{\tilde{\varphi}} + A_{12}A\sin\tilde{\varphi} + A_{12}B\sin\tilde{\tilde{\varphi}} - \\ \qquad\qquad -A_1 A\gamma_0\sin\tilde{\varphi} + RA_1\sin\varphi_1 = 0 \\ \dfrac{1}{\lambda}\alpha_2 B\sin\tilde{\tilde{\varphi}} + \dfrac{1}{\lambda}\alpha'_{12}A\sin\tilde{\varphi} + A_{12}A\sin\tilde{\varphi} + A_{12}B\sin\tilde{\tilde{\varphi}} - \\ \qquad\qquad -A_2 B\gamma_0\sin\tilde{\tilde{\varphi}} + RA_2\sin\varphi_1 = 0 \\ k_{0y}\alpha A_0 - k_{0y}\alpha R\cos\varphi_1 - \dfrac{1}{\gamma_0}A_1 R\sin\varphi_1 + AA_1\sin\tilde{\varphi} + \\ \qquad +\dfrac{1}{\gamma_0}A_2 R\sin\varphi_1 + BA_2\sin\tilde{\tilde{\varphi}} = 0 \quad (8)\end{cases}$$

The general solution of the system of equation (8) has the following form:

$$\begin{cases} A = \dfrac{2A_0 \gamma_0 k_{0y} \alpha \lambda (A_1 Y + A_2)}{(\alpha'_{12} + \lambda A_{12})(XY-1)\sqrt{(\gamma_0 k_{0y} \alpha)^2 + Z^2}} \\ B = \dfrac{2A_0 \gamma_0 k_{0y} \alpha \lambda (A_1 + A_2 X)}{(\alpha'_{12} + \lambda A_{12})(XY-1)\sqrt{(\gamma_0 k_{0y} \alpha)^2 + Z^2}} \\ R = A_0 \\ tg\tilde{\varphi} = \dfrac{Z}{\gamma_0 k_{0y} \alpha} \\ tg\tilde{\tilde{\varphi}} = \dfrac{Z}{\gamma_0 k_{0y} \alpha} \\ tg\varphi_1 = \dfrac{2\gamma_0 k_{0y} \alpha Z}{(\gamma_0 k_{0y} \alpha)^2 - Z^2} \end{cases}, \quad (9)$$

where the following expressions were used to reduce the solution:

$$X = \frac{\lambda \gamma_0 A_1 - \lambda A_{12} - \alpha_1}{\alpha'_{12} + \lambda A_{12}}, \quad Y = \frac{\lambda \gamma_0 A_2 - \lambda A_{12} - \alpha_2}{\alpha'_{12} - \lambda A_{12}},$$

$$Z = A_1 - A_2 - \frac{\lambda \gamma_0 (A_1(A_1 Y + A_2) + A_2(A_1 + A_2 X))}{(\alpha'_{12} + \lambda A_{12})(XY-1)}.$$

## 3 Results and Discussion

As it is well-known, the ratio between the frequency $\omega_s$ and the wave vector $\mathbf{k}$ of the spin wave – dispersion equation – determines the spectrum of spin waves in FM [20] and has the following form:

$$\omega_s(\mathbf{k}) = [\Omega_1 \Omega_2 + \\ + 4\pi g M_0 (\Omega_1 \cos^2 \varphi_k + \Omega_2 \sin^2 \varphi_k) \sin^2 \theta_k]^{1/2}. \quad (10)$$

where $\Omega_1$, $\Omega_2$ are frequencies of spin waves, $\theta_k$ and $\varphi_k$ are polar and azimuthal angles of the wave vector $\mathbf{k}$, respectively.

Assuming that FM has anisotropy of the "easy axis" ($\beta > 0$) and $\mathbf{M}_0 \parallel \mathbf{n} \parallel \mathbf{H}_0^{(i)}$, where $\mathbf{H}_0^{(i)}$ is internal magnetic field, then:

$$\Omega_1 = \Omega_2 = \Omega = g M_0 \left( \alpha k^2 + \frac{H_0^{(i)}}{M_0} + \beta \right). \quad (11)$$

In the AFMs, unlike the FMs, there are not one but two branches of spin waves. Considering that AFM has anisotropy of the "easy axis" (($\beta - \beta'$)>0) the frequencies of the spin waves are determined by the formulas:

$$\omega_{s1}(\mathbf{k}) = \Omega_+ = g M_0 \left[ 2\delta(\alpha - \alpha') k^2 + \left( \frac{H_1}{M_0} \right)^2 \right]^{1/2} + g H_0^{(e)}$$

$$\omega_{s2}(\mathbf{k}) = \Omega_- = g M_0 \left[ 2\delta(\alpha - \alpha') k^2 + \left( \frac{H_1}{M_0} \right)^2 \right]^{1/2} - g H_0^{(e)}$$

(12)

where $\mathbf{H}_0^{(e)}$ is external magnetic field and a magnetic field $H_1 = M_0 \sqrt{2\delta(\beta - \beta')}$.

Taking into account that components of the wave vector $\mathbf{k} = (k_{0x}, k_{0y}, k_{0z})$ for FM of dispersion equation can be presented as follows:

$$\begin{aligned} k_{0x} &= k_0 \sin \theta_k \cos \varphi_k \\ k_{0y} &= k_0 \sin \theta_k \sin \varphi_k \\ k_{0z} &= k_0 \cos \theta_k \end{aligned}, \quad (13)$$

where $\cos \varphi_k = 0$, $\sin \varphi_k = 1$ and $\theta_k = \theta + \dfrac{\pi}{2}$, then substituting Eqs. (11) and (13) in the dispersion equation for FM (10) $k_0$ can be expressed as:

$$k_0 = \frac{1}{\alpha} \sqrt{-\frac{H_0^{(i)}}{M_0} - \beta - 2\pi \cos \theta \pm \sqrt{4\pi^2 \cos^4 \theta + \left( \frac{\omega}{g M_0} \right)^2}}. \quad (14)$$

From dispersion equation for AFM (12) can be expressed the length of the spin wave $\lambda_{sw}$, taking into account that components of the wave vector $\mathbf{k} = \left( 0, \dfrac{i}{\lambda_{sw}}, k_{2z} \right)$:

$$\lambda_{sw\pm} = \sqrt{\frac{2\delta(\alpha - \alpha')}{\sqrt{2\delta(k_0^2 \sin^2 \theta (\alpha - \alpha') - (\beta - \beta')) - \left( \dfrac{\omega \pm g H_0^{(e)}}{g M_0} \right)^2}}}. \quad (15)$$

Using Eqs. (14) and (15) for the solution (9), the dependences of all parameters of surface evanescent spin wave in AFM when spin wave in FM falls onto the FM/AFM interface, namely $A$, $B$, $R$, $\tilde{\varphi}$, $\tilde{\tilde{\varphi}}$ and $\varphi_1$ on the frequency can be determined.

## 4 Conclusion

The abridged general boundary conditions in the interface between FM/AFM were considered in this work including the energies of uniform and non-uniform exchange between all sublattices based on the previous investigation [1] and the surface evanescent spin wave in form (7) transition through the FM/AFM interface was theoretically investigated in the case when the spin wave in FM falls onto this interface and is considered in AFM. The coefficients of transmission and reflection of surface evanescent spin wave and relevant phases were derived in form (9). The dependences of all parameters of surface evanescent spin wave were determined using FM and AFM dispersion equations.

As can be seen from the solution (9) the phases of transmitted spin wave for each sublattice of two-sublattice AFM are equal. Also it is obvious that the coefficient of reflection is same as the coefficient of incidence of spin wave $R = A_0$, which is correct, because otherwise it would contradict the continuity of the energy flow. Indeed, the wave in AFM is evanescent, and the energy is not transferred to the plus infinity.

**Acknowledgement**

This work was supported by the European Union's Horizon 2020 research and innovation programme under the Marie Sklodowska-Curie GA No. 644348 (MagIC).